\documentclass[prb,twocolumn,amsmath,amssymb,showpacs]{revtex4}

\usepackage{graphicx}
\usepackage{amssymb}

\newcommand{\be}{\beta}

\newcommand{\de}{\delta}

\newcommand{\pd}{\partial}

\newcommand{\str}{\text{Str}}

\def\be{\begin{equation}}
\def\ee{\end{equation}}
\def\bea{\begin{eqnarray}}
\def\eea{\end{eqnarray}}

\begin{document}
\title{Boundary multifractality in critical 1D systems with long-range
  hopping}

\author{A.~Mildenberger$^{1}$, A. R.~Subramaniam$^2$,
  R.~Narayanan$^{3,5}$,
F.~Evers$^{3,4}$, I. A.~Gruzberg$^2$, and
A.D.~Mirlin$^{3,4,*}$}

\affiliation{ $\mbox{}^{1}$ Fakult\"at f\"ur Physik, Universit\"at
Karlsruhe,
76128 Karlsruhe, Germany\\
\hbox{$^{2}$James Franck Institute,
University of Chicago, 5640 S. Ellis Ave., Chicago, IL 60637}\\
$\mbox{}^{3}$ Institut f\"ur Nanotechnologie, Forschungszentrum
Karlsruhe, 76021 Karlsruhe, Germany \\
$\mbox{}^{4}$ Inst. f\"ur Theorie der Kondensierten Materie,
Universit\"at Karlsruhe, 76128 Karlsruhe, Germany \\
$\mbox{}^{5}$ Department of Physics, Indian
Institute of Technology Madras, Chennai 600036, India
}

\date{\today}

\begin{abstract}
Boundary multifractality of electronic wave functions is studied
analytically and numerically for the power-law random banded matrix
(PRBM) model, describing a critical one-dimensional system with
long-range hopping. The peculiarity of the Anderson localization
transition in this model is the existence of a line of fixed points
describing the critical system in the bulk. We demonstrate that the
boundary critical theory of the PRBM model is not uniquely
determined by the bulk properties. Instead, the boundary criticality
is controlled by an additional parameter characterizing the hopping
amplitudes of particles reflected by the boundary.
\end{abstract}



\maketitle

\section{Introduction}
\label{s1}

Although almost half a century has passed after Anderson's seminal
paper \cite{anderson58}, the properties of disordered
systems at Anderson localization transitions remain a subject of
active current research. This interest is additionally
motivated by the understanding that, besides the
``conventional'' Anderson transition in $d>2$ dimensions, disordered
fermions in two dimensions (2D) possess a rich variety of critical
points governing quantum phase transitions in these systems.

One of the striking peculiarities of the Anderson transitions is the
multifractality of the electronic wave functions
\cite{Wegner80,Castellani-Peliti-86}, see
Refs.~\onlinecite{janssen,review} for recent reviews. Specifically,
the scaling of moments of the wave functions with the system size
$L$ is characterized  by a
continuum of independent critical exponents
$\tau_q$,
\begin{align}
L^d \langle |\psi({\bf r})|^{2q} \rangle &\sim L^{-\tau_q}, & \tau_q
&\equiv d(q-1) + \Delta_q, \label{e1b}
\end{align}
where $\langle\ldots\rangle$ denotes the disorder average. In the
field-theoretical ($\sigma$-model) language,\cite{Wegner80} $\tau_q$
are scaling dimensions of higher-order operators describing the
moments of the local density of states. Note that one often
introduces fractal dimensions $D_q$ via $\tau_q=D_q(q-1)$. In a
metal $D_q=d$, while at
a critical point $D_q$ is a nontrivial
function of $q$, implying the multifractality of wave functions.
Nonvanishing anomalous dimensions $\Delta_q\equiv (q-1)(D_q-d)$
distinguish a critical point from a metallic phase and determine the
scaling of wave function correlations. Among them, $\Delta_2 < 0$
plays the most prominent role, governing the spatial correlations of
the intensity $|\psi|^2$,
\begin{equation}
\label{e2} L^{2d} \langle |\psi^2 ({\bf r}) \psi^2({\bf r}')|\rangle
\sim (|{\bf r} - {\bf r}'|/L)^{\Delta_2}.
\end{equation}
This equation, which in technical terms results from an operator
product expansion of the field theory \cite{duplantier91}, can be
obtained from (\ref{e1b}) by using the fact that the wave function
amplitudes become essentially uncorrelated at $|{\bf r} - {\bf
r}'|\sim L$. Scaling behavior of higher order spatial correlations,
$\langle|\psi^{2q_1}({\bf r}_1) \psi^{2q_2}({\bf r}_2) \ldots
\psi^{2q_n}({\bf r}_n)|\rangle$, can be found in a similar way.
Above, the points ${\bf r}_i$ were assumed to lie in the bulk of a
critical system. In this case we denote the multifractal exponents
by $\tau_q^{\rm b}$, $\Delta_q^{\rm b}$, etc. The multifractality of
wave functions has been studied analytically and numerically for a
variety of systems: Anderson transition in $d = 2 + \epsilon$, 3,
and 4 dimensions \cite{Wegner80,akl,MF-Anderson-numerics}, as well
as weak multifractality \cite{MF-weak}, Dirac fermions in random
gauge fields \cite{MF-Dirac}, symplectic-class Anderson transition
\cite{MF-symplectic}, integer quantum Hall \cite{MF-IQH} and
spin quantum Hall \cite{Mirlin03} transitions in two dimensions, and
power-law random banded matrices \cite{MF-RBM}.

Recently, the concept of the wave function multifractality was
extended \cite{subramaniam06} to the surface of a system at the
critical  point of an Anderson transition. It was shown that the
fluctuations of critical wave functions at the surface are
characterized by a new set of exponent $\tau_q^{\rm s}$ (or,
equivalently, anomalous exponents $\Delta_q^{\rm s})$, which are in
general independent from their bulk counterparts. This boundary
critical behavior was explicitly studied, analytically as well as
numerically, for the 2D spin quantum Hall transition
\cite{subramaniam06,sqhe-unpub} and a 2D weakly localized metal
\cite{subramaniam06} and, most recently, for the Anderson transition
in a 2D system with spin-orbit coupling \cite{obuse06}.

In the present paper, we analyze the boundary criticality in the
framework of the power-law random banded matrix (PRBM) model
\cite{prbm,review,MF-RBM}. The model is defined \cite{prbm} as the
ensemble of random Hermitian matrices $\hat H$ (real for $\beta=1$
or complex for $\beta=2$). The matrix elements $H_{ij}$ are
independently distributed Gaussian variables with zero mean $\langle
H_{ij}\rangle=0$ and the variance
\begin{equation}
\label{e4} \langle |H_{ij}|^2\rangle =a^2(|i-j|)\ ,
\end{equation}
where $a(r)$ is given by
\begin{equation}
a^2(r)={1\over 1+(r/b)^{2\alpha}}\ . \label{e6}
\end{equation}
At $\alpha=1$ the model undergoes an Anderson transition from the
localized ($\alpha>1$) to the delocalized ($\alpha<1$) phase. We
concentrate below on the critical value $\alpha=1$, when $a(r)$
falls down as $a(r)\propto 1/r$ at $r\gg b$.

In a straightforward interpretation, the PRBM model describes a 1D
sample with random long-range hopping, the hopping amplitude
decaying as $1/r^\alpha$ with the distance. Also, such an ensemble
arises as an effective description in a number of physical contexts
(see Ref.~\onlinecite{MF-RBM} for relevant references). At
$\alpha=1$ the PRBM model is critical for arbitrary values
of $b$ and
shows all the key features of an 
Anderson critical point, including
multifractality of eigenfunctions and nontrivial spectral
compressibility \cite{prbm,review}. The existence of the parameter
$b$ which labels critical points is a distinct feature of the
PRBM model: Eq.~(\ref{e4}) defines a whole family of critical
theories parametrized by $b$.  The limit $b\gg 1$ represents a
regime of weak multifractality, analogous to the conventional
Anderson transition in $d=2+\epsilon$ with $\epsilon\ll 1$. This
limit allows for a systematic analytical treatment via the mapping
onto a supermatrix $\sigma$-model and the weak-coupling expansion
\cite{prbm,review,MF-RBM}. The opposite limit $b\ll 1$ is
characterized by very strongly fluctuating eigenfunctions, similarly
to the Anderson transition in $d\gg 1$, where the transition takes
place in the strong disorder (strong coupling in the
field-theoretical language) regime. It is also accessible to an
analytical treatment using a real-space renormalization-group (RG)
method \cite{MF-RBM} introduced earlier for related models in
Ref.~\onlinecite{levitov90}.

In addition to the feasibility of the systematic analytical
treatment of both the weak-coupling and strong-coupling regimes, the
PRBM model is very well suited for direct numerical simulations in a
broad range of couplings. For these reasons, it has attracted a
considerable interest in the last few years as a model for the
investigation of various properties of the Anderson critical point
\cite{MF-RBM,prbm-papers}. We thus employ the PRBM model for the
analysis of the boundary multifractality in this work. The existence
of a line of fixed points describing the critical system in the bulk
makes this problem particularly interesting. We will demonstrate
that the boundary critical theory of the PRBM model is not uniquely
determined by the bulk properties. Instead, the boundary criticality
is controlled by an additional parameter characterizing the hopping
amplitudes of particles reflected by the boundary.

The structure of the paper is as follows. In Sec.~\ref{s2} we
formulate the model. Section \ref{s3} is devoted to the analytical
study of the boundary multifractal spectrum, with the two limits
$b\gg 1$ and $b\ll 1$ considered in Secs.~\ref{s3.1} and \ref{s3.2},
respectively. The results of numerical simulations are presented in
Sec.~\ref{s4}. Section \ref{s5} summarizes our findings.

\section{Model}
\label{s2}

We consider now the critical PRBM model with a boundary at
$i=0$, which means that the matrix element $H_{ij}$ is zero whenever
one of the indices is negative. The important point is that, for a
given value of the bulk parameter $b$,  the implementation of the
boundary is not unique, and that this degree of freedom will affect
the boundary criticality. Specifically, we should specify what
happens with a particle which ``attempts to hop'' from a site $i\ge
0$ to a site $j<0$, which is not allowed due to the boundary. One
possibility is that such hops are simply discarded, so that the
matrix element variance is simply given by $\langle
|H_{ij}|^2\rangle = [1+(i-j)^2/b^2]^{-1}$ for $i,j\ge 0$. More
generally, the particle may be reflected by the boundary with
certain probability $p$ and ``land'' on the site $-j>0$. This leads
us to the following formulation of the model:
\begin{eqnarray}
\label{e4a}
&& \langle |H_{ij}|^2\rangle =J_{ij}\ , \\
\label{e7} && J_{ij} = \dfrac{1}{1+ \left| i-j \right|^2/b^2} +
\dfrac{p}{1+ \left| i+j \right|^2/ b^2}.
\end{eqnarray}
While the above probability interpretation restricts $p$ to the
interval $[0,1]$, the model is defined for all $p$ in the range
$-1 < p < \infty$. The newly introduced parameter $p$ is immaterial
in the bulk, where $i,j \gg |i-j|$ and the second term in
Eq.~(\ref{e7}) can be neglected. Therefore, the bulk exponents
$\tau_q^{\rm b}$ depend on $b$ only (and not on $p$), and their
analysis performed in Ref.~\onlinecite{MF-RBM} remains applicable
without changes. On the other hand, as we show below by both
analytical and numerical means, the surface exponents 
$\tau_q^{\rm s}$ are a function of two parameters, $b$ and $p$.

Equation (\ref{e7}) describes a semi-infinite system with one
boundary at $i=0$. For a finite system of a length $L$ (implying
that the relevant coordinates are restricted to $0\le i,j \le L$)
another boundary term, $p'/[1+ (i+j-2L)^2 /b^2]$, is to be included
on the right-hand side of Eq.~(\ref{e7}). In general, the parameter
$p'$ of this term may be different from $p$. This term, however,
will not affect the boundary criticality at the $i=0$ boundary, so
we discard it below.

\section{Boundary mutlifractality: Analytical methods}
\label{s3}

\subsection{$b\gg 1$}
\label{s3.1}

The regime of weak criticality, $b\gg 1$, can be studied via a
mapping onto the supermatrix $\sigma$-model
\cite{prbm,review,MF-RBM}, in analogy with the conventional random
banded matrix model \cite{fyodorov94}. The $\sigma$-model action has
the form
\begin{equation}
\label{e8} S[Q] = {\beta\over 4}\str \left[ (\pi \nu)^2
\sum_{i,j=0}^{\infty} J_{ij} Q_i Q_j - i \pi \nu \omega
\sum_{i=0}^{\infty} Q_i \Lambda \right],
\end{equation}
where  $Q_r$ is a $4\times 4$ ($\beta=2$) or $8\times 8$ ($\beta=1$)
supermatrix field constrained by $Q^2_r=1$, $\Lambda = {\rm
diag}({\bf 1},-{\bf 1})$, and $\str$ denotes the
supertrace.\cite{efetov-book} Furthermore, $J_{ij}$ are given by
Eq.~(\ref{e7}), $\omega$ is the frequency, and $\nu$ is the density
of states given by the Wigner semicircle law
\begin{equation}
\label{e9} \nu(E)={1\over2\pi^2 b} (4\pi b - E^2)^{1/2}\ , \qquad
|E|<2\sqrt{\pi b}.
\end{equation}
For definiteness, we will restrict ourselves to the band center,
$E=0$, below.

To calculate the multifractal spectrum to the leading order in
$1/b\ll 1$, we will need the quadratic form of the action (\ref{e8})
expressed in terms of independent coordinates. Parametrizing the
field $Q$ (constrained to $Q^2=1$) in the usual way,
\begin{equation}
\label{e10} Q_i = \Lambda \left(1 + W_i + \frac{W_i^2}{2} +
\ldots\right),
\end{equation}
we obtain the action to the second order in the $W$ fields,
\begin{equation}
\label{e11} S[W] = {\pi\nu\beta\over 4} \str \sum_{i,j=0}^{\infty}
W_i \left[2\pi\nu (J_0^{(i)}
\de_{ij}-J_{ij})-i\omega\delta_{ij}\right]W_j,
\end{equation}
where
\begin{equation}
\label{e12} J_0^{(i)} = \sum_{k=0}^{\infty}J_{ik}.
\end{equation}
The equation of motion for this action reads (after the Fourier
transformation from the frequency into the time domain)
\begin{equation}
\label{e13} \frac{\pd W_i(t)}{\pd t} + \pi \nu \sum_{j=0}^{\infty}
\left[\de_{ij} J_0^{(i)} - J_{ij}  \right] W_j(t) = 0.
\end{equation}
This equation is the analog of the diffusion equation for a metallic
system.

The $\sigma$-model action allows us to calculate the moments
$\langle|\psi^2_r|^q\rangle$ at a given point $r$. On the
perturbative level, the result reads
\cite{fyodorov94,prbm,review,MF-RBM}
\begin{equation}
\langle|\psi^2_r|^q\rangle = {\langle|\psi^2_r|^q\rangle}_{\rm RMT}
\left[1+{1\over\beta}q(q-1)\Pi_{rr}\right]. \label{e14}
\end{equation}
Here the factor ${\langle|\psi^2_r|^q\rangle}_{\rm RMT}$ is the
random-matrix-theory result equal to $(2q-1)!!L^{-q}$ for $\beta=1$
and $q!L^{-q}$ for $\beta=2$. The second term in the square brackets
in Eq.~(\ref{e14}), which constitutes the leading perturbative
correction, is governed by the return probability $\Pi_{rr}$ to the
point $r$, i.e., the diagonal matrix element of the generalized
diffusion propagator $\Pi_{rr'}$. The latter is obtained by the
inversion of the kinetic operator of Eqs.~(\ref{e11}), (\ref{e13}),
\begin{equation}
\label{e15} \pi \nu \sum_{j=0}^{\infty} \left[\de_{ij} J_0^{(i)} -
J_{ij}  \right] \Pi_{jm} = \delta_{im}- L^{-1}.
\end{equation}
(The ``diffusion'' operator has a zero mode related to the
particle conservation. The term $L^{-1}$ on the right-hand side 
of Eq. (\ref{e15})
ensures that the inversion is taken on the subspace of nonzero modes.)
In the bulk case, the inversion is easily performed via the
Fourier transform,
\begin{equation}
\label{e16} \Pi_{rr'} \longrightarrow \tilde{\Pi}(k) = {t\over 8
|k|},  \qquad |k| \ll b^{-1},
\end{equation}
with $t^{-1} = {\pi\over 4}(\pi\nu)^2b^2$, i.e. $t=4/b$ at the band
center. The $1/|k|$ behavior of the propagator should be contrasted
to its $1/k^2$ scaling for a conventional metallic (diffusive)
system. This implies that the kinetics governed by Eq.~(\ref{e13})
is superdiffusive, also known as L\'evy flights \cite{levy-flights}.
Substitution of (\ref{e16}) in Eq.~(\ref{e14}) yields  a logarithmic
correction to the moments of the wave function amplitude,
\begin{equation}
\langle|\psi^2_r|^q\rangle = {\langle|\psi^2_r|^q\rangle}_{\rm RMT}
\left[1+ {q(q-1) \over 2\pi\beta b} \ln {L\over b}\right].
\label{e17}
\end{equation}
Equation (\ref{e17}) is valid as long as the relative correction is
small. The logarithmic divergence of the return probability in the
limit $L\to \infty$, which is a signature of criticality, makes the
perturbative calculation insufficient for large enough $L$. The
problem can be solved then by using the renormalization group (RG),
\cite{prbm,review,MF-RBM} which leads to the exponentiation of
the perturbative correction in
Eq.~(\ref{e17}). This results in Eq.~(\ref{e1b}) with the bulk
multifractal exponents
\begin{equation}
\label{e18} \tau_q^{\rm b} = (q-1)\left(1- {q\over 2\pi\beta
b}\right).
\end{equation}
The first term (unity) in the second factor in Eq.~(\ref{e18})
corresponds to the normal (metallic) scaling, the second one
determines the anomalous exponents
\begin{equation}
\label{e19} \Delta_q^{\rm b} = {q(1-q) \over 2\pi\beta b}.
\end{equation}
At the boundary, the behavior is qualitatively the same: the return
probability $\Pi_{rr}$ increases logarithmically with the system
size $L$, in view of criticality. However, as we show below, the
corresponding prefactor [and thus the prefactor in front of the
second term in square brackets in Eq.~(\ref{e17})] is different.
After the application of the RG this prefactor emerges in the
anomalous exponent,
\begin{equation}
\label{e20} \Delta_q^{\rm s} = {q(1-q) \over 2\pi\beta b} R_p \equiv
\Delta_q^{\rm b} R_p.
\end{equation}
In the presence of a
boundary the system is not translationally
invariant anymore, which poses an obstacle for an analytical
calculation of the return probability $\Pi_{rr}$. While for
L\'evy-flight models with absorbing boundary (that is obtained from
our equation (\ref{e13}) with $p=0$ by a replacement of $J_0^{(i)}$
with its  bulk value $J_0$) an analytical progress can be achieved
via the Wiener-Hopf method \cite{zumofen95}, it is not applicable in
the present case, since the kernel of Eq.~(\ref{e13}) is not a
function of $i-j$ only. We thus proceeded by solving the classical
evolution equation (\ref{e13}) numerically with the initial
condition $W_i(0) = \delta_{ir}$. The value $W_r(t)$ of the solution
at the point $r$ (i.e. the probability to find the particle at the
initial point) decays with the time as $1/t$, so that the integral
$\int dt W_r(t)$ yields the logarithmically divergent return
probability discussed above. Extracting the corresponding prefactor,
we find the anomalous exponent,
\begin{equation}
\label{e21} {\Delta_q\over q(1-q)} = {1\over\beta} t W_{r}(t)
|_{t\to\infty}.
\end{equation}
Note that the limit of the large system size $L\to\infty$ should be
taken in Eq.~(\ref{e21}) before $t\to\infty$, so that the particle
does not reach the boundary for $r$ in the bulk (or, for $r$ at the
boundary, does not reach the opposite boundary).

\begin{figure}
\begin{center}
\includegraphics[width=0.95\columnwidth,clip]{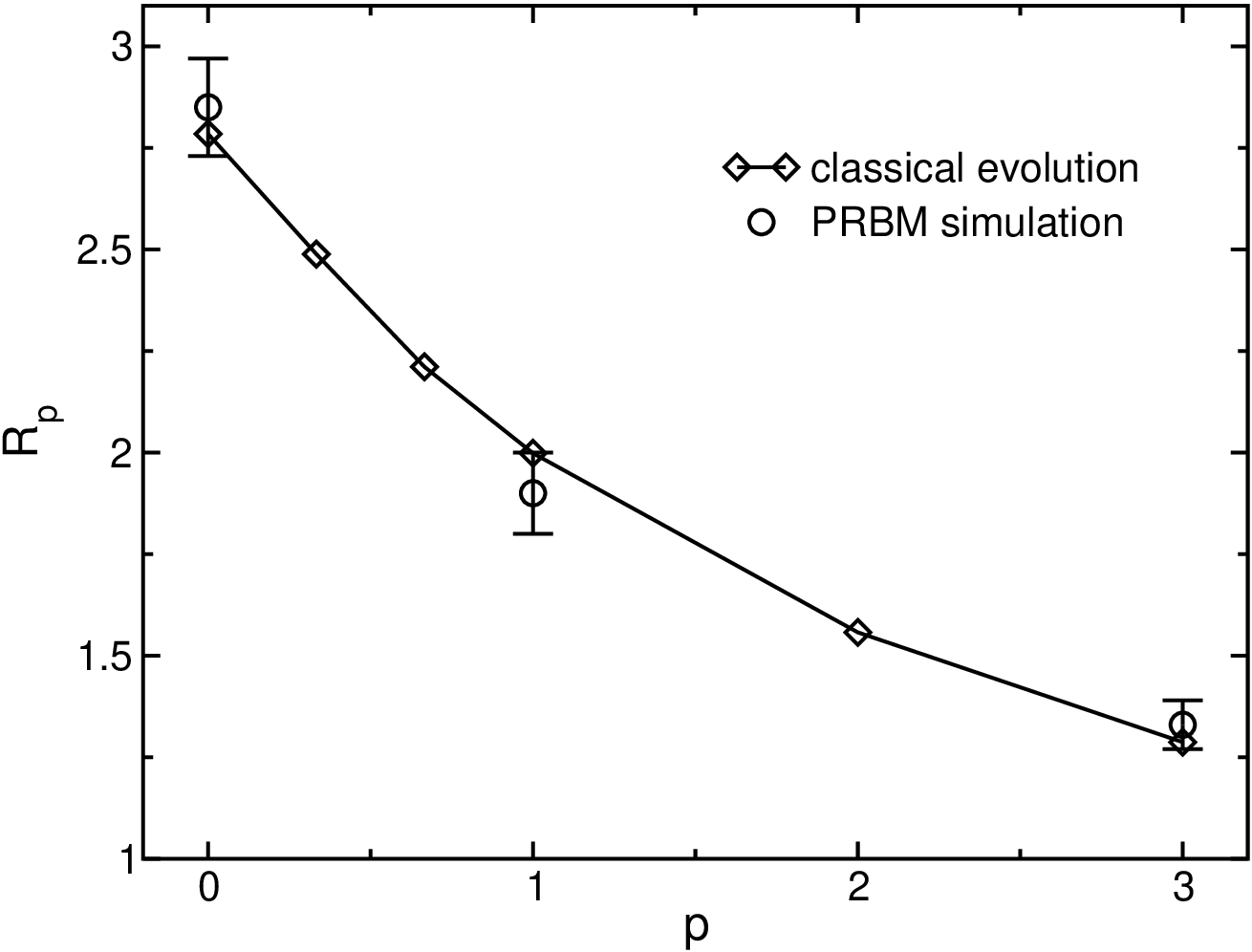}
\caption{The ratio $R_p = \Delta_q^{\rm s}(b,p) / \Delta_q^{\rm
b}(b)$ of the surface and bulk anomalous exponents for
large $b$, as
a function of the reflection parameter $p$. Diamonds represent the
results of the $\sigma$-model analysis with a numerical solution of
the corresponding classical evolution equation, as described in
Sec.~\ref{s3.1}. Circles represent a direct computer simulation of
the PRBM model, Eqs.~(\ref{e4a}), (\ref{e7}), see Sec.~\ref{s4},
with $b=8$. The ratio $R_p$ has been evaluated for the range $0 < q
< 1$, where the numerical accuracy of the anomalous exponents is the
best. Within this interval we find that $R_p$ is $q$-independent
(within numerical errors) in agreement with Eq.~(\ref{e20}). }
\label{fig1}
\end{center}
\end{figure}

We have checked that the numerical implementation of
Eqs.~(\ref{e21}), (\ref{e13}) reproduces the analytical result
(\ref{e19}) in the bulk. We then proceeded with numerical evaluation
of the surface multifractal exponents $\Delta_q^{\rm s}$. For this
purpose, we have discretized the time variable in Eq.~(\ref{e13})
with a step $\Delta t = 1/2$. With the parameters $\pi\nu= 1/\pi$,
$b=10$, $L=10000$, and $t=500$, the product $tW_r(t)$ yields its
required asymptotic value with the accuracy of the order of $2\%$.
The results for the corresponding prefactor $R_p$, as defined in
Eq.~(\ref{e20}), are shown in Fig.~\ref{fig1} for several values of
$p$ between 0 and 3. It is seen that the boundary exponents not only
differ from their bulk counterparts but also depend on $p$.

For the particular case of the reflection probability $p=1$ we can
solve the evolution equation (\ref{e13}) and find $\Delta_q^{\rm s}$
analytically. Indeed, the corresponding equation can be obtained from
its bulk counterpart (defined on the whole axis, $-\infty < i
<\infty$) by ``folding the system'' on the semiaxis $i>0$ according
to $W_i(t)+W_{-i}(t) \longrightarrow W_i(t)$, cf.
Ref.~\onlinecite{levy-p1}. This clearly leads to a doubling of the
return probability, so that
\begin{equation}
\label{e22} R_1 = 2,
\end{equation}
in full agreement with the numerical solution of the evolution
equation.

\subsection{$b\ll 1$}
\label{s3.2}

In the regime of small $b$ the eigenstates are very sparse. In this
situation, the problem can be studied by a real-space RG method that
was developed in Ref.~\onlinecite{levitov90} for related models and
in Ref.~\onlinecite{MF-RBM} for the PRBM model. Within this
approach, one starts from the diagonal part of the Hamiltonian and
then consecutively includes into consideration nondiagonal matrix
elements $H_{ij}$ with increasing distance $\rho=|i-j|$. The central
idea is that only rare events of resonances between pairs of remote
states are important,  and that there is an exponential hierarchy of
scales at which any given state finds a  resonance partner:
$\ln\rho_1 \sim b^{-1}$, \, $\ln\rho_2\sim b^{-2}$, \ldots. This
allows one to formulate RG equations for evolution of quantities of
interest with the ``RG time'' $t=\ln\rho$. We refer the reader for
technical details of the derivation to Ref.~\onlinecite{MF-RBM}
where the evolution equation of the distribution $f(P_q)$ of the
inverse participation ratios, $P_q=\sum_r|\psi_r^2|^q$, as well as
of the energy level correlation function, was derived. In the
present case, we are interested in the statistics of the local
quantity, the wave function intensity $|\psi_r^2|$ at a certain point
$r$. Assuming first that $r$ is in the bulk and generalizing the
derivation of Ref.~\onlinecite{MF-RBM}, we get the evolution
equation for the corresponding distribution function
$f(y\equiv|\psi_r^2|)$,
\begin{eqnarray}
\label{e23} {\partial f(y,\rho)\over\partial\ln\rho} &=&
{2b\over\pi} \int_0^{\pi/2}{d\theta\over\sin^2\theta\cos^2\theta}
\int_{-\infty}^\infty dy' f(y',\rho)   \nonumber \\
&\times& [\delta(y-y'\cos^2\theta) + \delta(y-y'\sin^2\theta)
\nonumber \\
&& - \delta(y-y') -\delta(y)].
\end{eqnarray}
Equation (\ref{e23}) is written for $\beta=1$; in the case of
$\beta=2$ one should make a replacement $b\longrightarrow
(\pi/2\sqrt{2})b$.
The physical meaning of Eq.~(\ref{e23}) is rather
transparent: its right-hand side is a ``collision integral''
describing a resonant mixture of two states with the intensities
$y'$ and $0$ at the point $r$, leading to formation of superposition
states with the intensities $y'\cos^2\theta$ and  $y'\sin^2\theta$.
Multiplying Eq.~(\ref{e23}) by $y^q$ and integrating over $y$, we
get the evolution equation for the moments $\langle y^q\rangle$,
\begin{equation}
\label{e24} {\partial \langle y^q\rangle \over\partial\ln\rho} = - 2
b T(q)\langle y^q\rangle \ ,
\end{equation}
where
\begin{eqnarray}
\label{e25} T(q) &=& {1\over\pi}
\int_0^{\pi/2}{d\theta\over\sin^2\theta\cos^2\theta}
(1-\cos^{2q}\theta - \sin^{2q}\theta)
\nonumber \\
&=& {1\over 2^{2q-3}} {\Gamma(2q-1) \over \Gamma(q) \Gamma(q-1)}.
\end{eqnarray}
The RG should be run until $\rho$ reaches the system size $L$. Thus,
the bulk multifractal exponents are equal to
\begin{equation}
\label{e26} \tau_q^b = 2bT(q),
\end{equation}
in agreement with Ref.~\onlinecite{MF-RBM}.

How will the evolution equation (\ref{e23}) be modified if the point
$r$ is located at the boundary? First, the factor 2 on the
right-hand side of (\ref{e23}) will be absent. Indeed, this factor
originated from the probability to encounter a resonance. In the
bulk, the resonance partner can be found either to the left or to
the right, thus the
factor of two. For a state at the boundary only one
of these possibilities remains, so this factor is absent. Second,
one should now take into account also the second term in the variance
$J_{ij}$ of the matrix element $H_{ij}$, Eq.~(\ref{e7}). In view of
the hierarchy of resonances described above, the relevant matrix
elements will always connect two points, one of which is much closer
to the boundary than the other (say, $i\ll j$). In this situation,
the two terms in (\ref{e7}) become equivalent
(up to the prefactor
$p$ in the second term) and can be combined,
\begin{equation}
\label{e27} J_{ij} \simeq {(1+p) b^2\over j^2}\ , \qquad i\ll j.
\end{equation}
Therefore, the effect of the second term amounts to the rescaling $b
\longrightarrow (1+p)^{1/2} b$. Combining both the effects, we get
the boundary multifractal exponents,
\begin{equation}
\label{e28} \tau_q^{\rm s} = (1+p)^{1/2} b T(q) = {(1+p)^{1/2} \over
2} \tau_q^{\rm b}.
\end{equation}
The above real-space RG method works for $q>1/2$, where the
multifractal exponent $\tau_q$ is small \cite{note1}. The results
can, however be extended to the range of $q<1/2$ by using the
recently found symmetry relation between the multifractal exponents
\cite{mirlin06},
\begin{equation}
\label{e29} \Delta_q = \Delta_{1-q}.
\end{equation}
Independently of whether $q$ is larger or smaller than 1/2, the
obtained relation between the surface and the bulk multifractal
spectra can be formulated in the following way:
\begin{equation}
\label{e30} \tau_q^{\rm s} (b,p) = \tau_q^{\rm b} (b \longrightarrow
b(1+p)^{1/2}/2).
\end{equation}

\section{Boundary multifractality: Numerical simulations}
\label{s4}

In this section we present the results for the multifractality
spectra obtained by direct numerical simulations of the PRBM model,
Eqs.~(\ref{e4a}), (\ref{e7}), with $\beta=1$. 
The model has been implemented with
two boundaries, each one having the same boundary parameter $p$.
Using standard diagonalization routines, systems with sizes
$L=128,\,256,\,512,\,1024,\, 2048$ and $4096$ sites have been
studied with ensembles comprising of $ 10^7$ ($L=128$, $3\cdot10^8$
wave functions) to $5000$ ($L=4096$, $5\cdot10^6$ wave functions)
disorder configurations. The multifractal analysis has been
performed with intensities $|\psi_r^2|$ averaged (coarse grained)
over blocks of four neighboring sites in order to access negative
$q$ values, $q \gtrsim -2$. For the analysis of the surface
multifractal exponents, only the four sites closest to boundaries
have been taken into account.

\begin{figure}
\begin{center}
\includegraphics[width=0.95\columnwidth,clip]{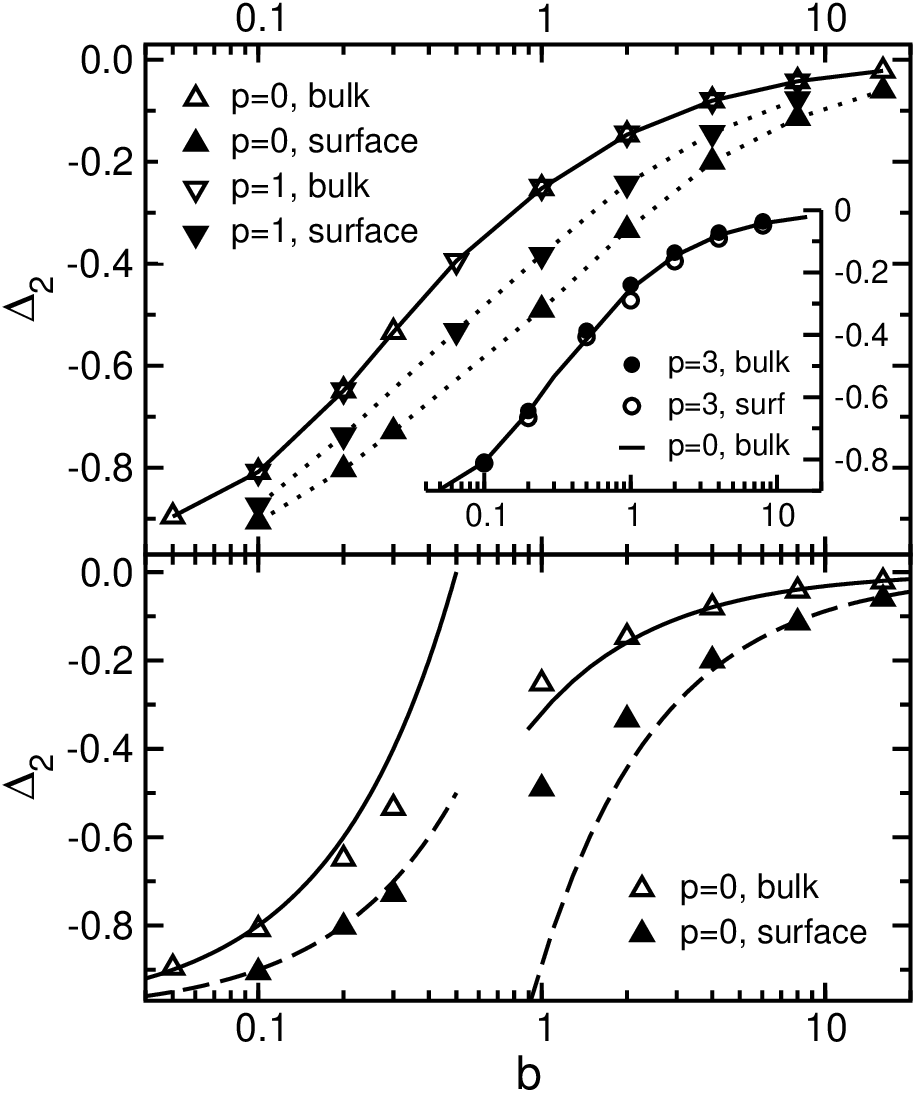}
\caption{Upper panel: Anomalous exponent $\Delta_2 \equiv D_2-1$ as
a function of $b$ from numerical simulations in the bulk and at the
boundary for the reflection parameter $p=0$ and $1$. The inset shows
data for $p=3$ compared to the $p=0$ bulk values. Lower panel:
Surface and bulk data for $p=0$ compared with analytical results for
small and large $b$ (using $R_0=2.78$), Eqs.~(\ref{e19}),
(\ref{e20}), (\ref{e26}), (\ref{e28}).} \label{fig2}
\end{center}
\end{figure}

Figure \ref{fig2} illustrates nicely our main findings. We show
there the dependence of the anomalous dimension $\Delta_2 \equiv D_2
-1$ on $b$ in the bulk and at the boundary, for three different
values of the reflection parameter $p$. It is seen, first of all,
that the bulk exponent $\Delta_2^b$ does not depend on $p$, in
agreement with the theory. Second, the boundary exponent
$\Delta_2^s$ is different from the bulk one. Third, the boundary
exponent is not determined by $b$ only,
but rather depends on the boundary
parameter $p$ as well. The lower panel of Fig.~\ref{fig2}
demonstrates the agreement between the numerical results and the
analytical asymptotics of small and large $b$.

\begin{figure}
\begin{center}
\includegraphics[width=0.95\columnwidth,clip]{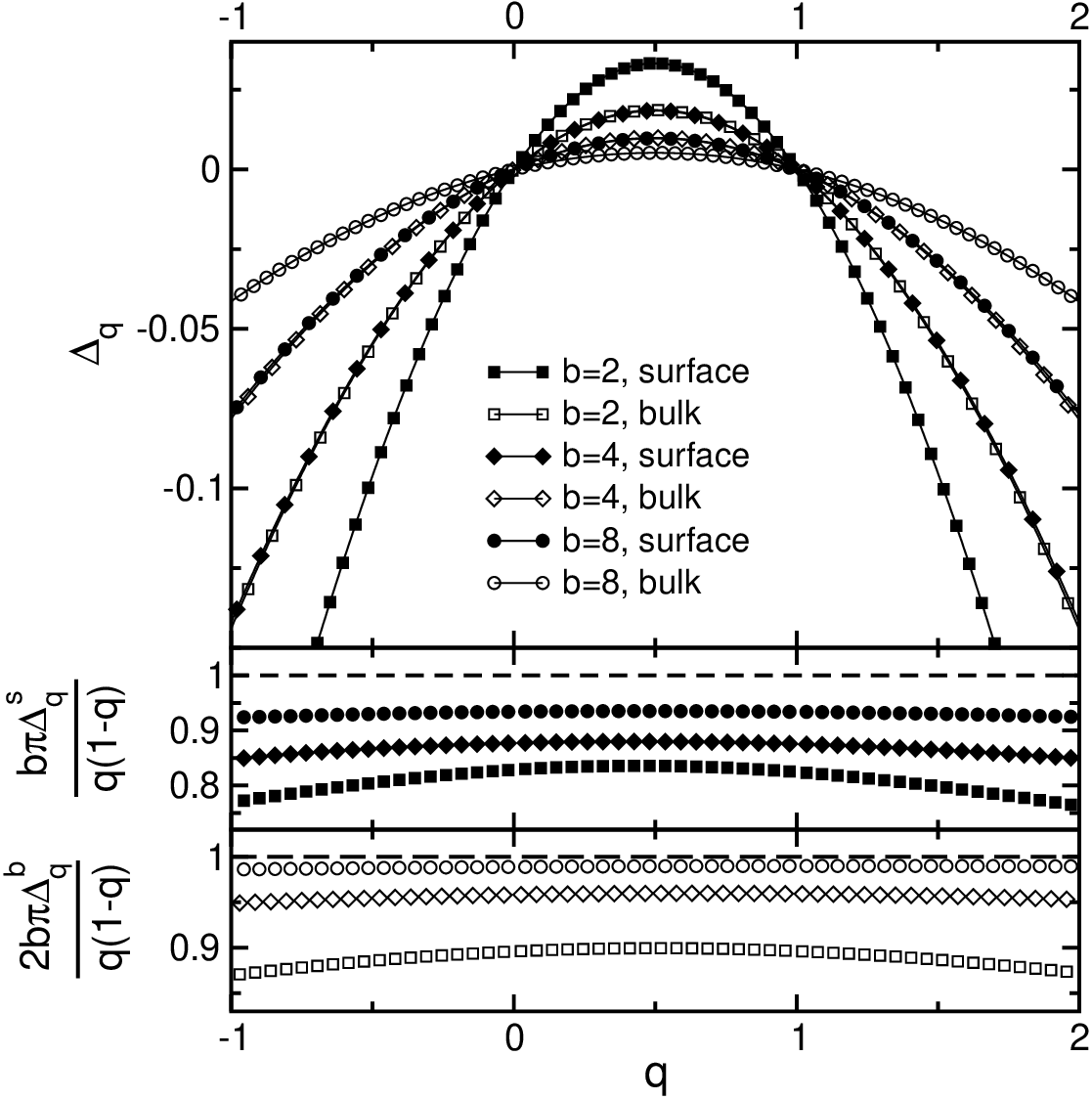}
\caption{Upper panel: Boundary and bulk
  multifractal spectra, $\Delta_q^{\rm s}$
  and $\Delta_q^{\rm b}$, at $b=2$, $4$, and $8$ for the
  reflection parameter $p=1$. In accordance with Eq.~(\ref{e22}),
  the surface multifractality spectrum is enhanced by a factor
  close to two compared to the bulk.\\
  Middle panel: Surface spectrum divided by the analytical
  large-$b$ result, Eq.~(\ref{e20}). The dashed line represents
  the analytical result for $b \gg 1$. With increasing $b$, the
  numerical data nicely converges towards the analytical result.\\
  Lower panel: Analogous plot for the bulk spectrum, Eq.~(\ref{e19}).
  The error estimate from the finite size extrapolation is
  $3\%$.
} \label{fig3}
\end{center}
\end{figure}

Having discussed the $b$-dependence of the fractal exponent with
fixed $q$ (equal to 2) shown in Fig.~\ref{fig2}, we turn to
Fig.~\ref{fig3}, where the whole multifractal spectra $\Delta_q$ are
shown for fixed large values of
$b$. Specifically, the anomalous dimensions
$\Delta_q^{\rm s}$ and $\Delta_q^{\rm b}$ are presented for
$b=2,\,4$, and 8, with the reflection parameter chosen to be $p=1$.
For all curves the $q$ dependence is approximately parabolic, as
predicted by the large-$b$ theory, Eqs.~(\ref{e19}) and (\ref{e20}),
with the prefactor inversely proportional to $b$. To clearly
demonstrate this, we plot in the lower two panels the exponents
$\Delta_q$ divided by the corresponding analytical results of the
large-$b$ limit. While for moderately large $b$ the ratio shows some
curvature, the latter disappears with increasing $b$ and the ratio
approaches unity, thus demonstrating the full agreement between the
numerical simulations and the analytical predictions. It is also
seen in Fig.~\ref{fig3} that the bulk multifractality spectrum for
$b=4$ and the surface spectrum for $b=8$ are almost identical, in
agreement with Eq.~(\ref{e22}).  The same is true for the relation
between the bulk spectrum for $b=2$ and the surface spectrum for
$b=4$.

We have further calculated the ratio of the large-$b$ surface and
bulk anomalous dimensions, $R_p = \Delta_q^{\rm s} / \Delta_q^{\rm
b}$, for several values of the reflection parameter, $p=0$, 1, and 3.
As shown in Fig.~\ref{fig1}, the results are in good agreement with
the $\sigma$-model predictions for $R_p$ obtained in
Sec.~\ref{s3.1}.

\begin{figure}
\begin{center}
\includegraphics[width=0.95\columnwidth,clip]{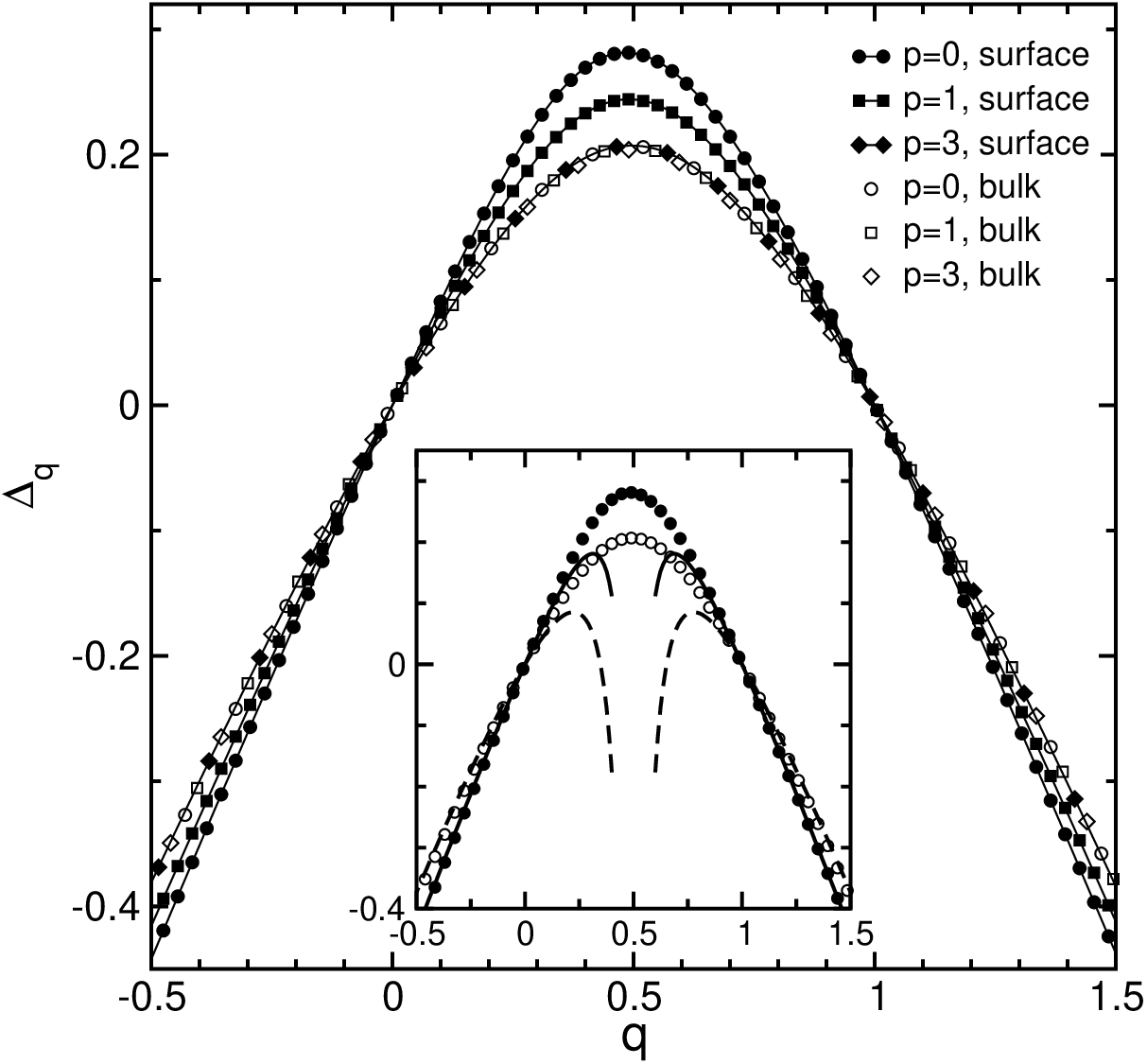}
\caption{Main panel: Numerically determined boundary and bulk
anomalous dimensions $\Delta_q$ at $b=0.1$ for $p=0$, $1$, and $3$.
As expected, the bulk anomalous dimension is independent of the
value of $p$. In accordance with Eq.~(\ref{e30}), for $p=3$ surface
and bulk dimensions have
the same values.\\
Inset: The $p=0$ data compared to the analytical results, surface
[solid line, Eq.~(\ref{e28})] and bulk [dashed line,
Eq.~(\ref{e26})]. Analytical data have been calculated for $q \geq
0.6$ and mirrored for $q \leq 0.4$ by using the symmetry relation
$\Delta_q = \Delta_{1-q}$. Note that the analytical result
(\ref{e26}) breaks down in the vicinity of $q=1/2$, at
$|q-1/2|\lesssim 1/\ln b^{-1}$, see Ref.~\onlinecite{note1}. 
At the boundary, one should replace
$b \to b(1+p)^{1/2}/2$ in this condition. We see, indeed, that for
$b=0.1$, $p=0$ the analytical formula works up to $|q-1/2|\simeq 0.4$
in the bulk and  $|q-1/2|\simeq 0.3$ at the boundary. } \label{fig4}
\end{center}
\end{figure}

In Fig.~\ref{fig4} the surface and bulk multifractal spectra are
shown for the case of small $b$. While the spectra are strongly
nonparabolic in this limit, they clearly exhibit the symmetry $q\to
1-q$, Eq.~(\ref{e29}). The data are in good agreement with the RG
results of Sec.~\ref{s3.2}. In particular, the surface spectrum for
$p=3$ is essentially identical to the bulk spectrum, as predicted by
Eq.~(\ref{e30}). In the inset, the surface and bulk multifractality
spectra for $p=0$ are compared with the analytical asymptotics,
Eqs.~(\ref{e26}), (\ref{e28}), supplemented by the symmetry relation
(\ref{e29}). Again, a very good agreement is seen, except for a
vicinity of $q=1/2$, where Eqs.~(\ref{e26}), (\ref{e28}) 
break down.\cite{note1}

\section{Conclusions}
\label{s5}

In summary, we have studied the boundary multifractality of wave
functions in the PRBM model describing a critical 1D system with
long-range hopping. Our findings strongly corroborate the ubiquity
of the notion of boundary mutlifractality (recently introduced in
Ref.~\onlinecite{subramaniam06}) in the context of disordered
electronic systems at criticality. We have demonstrated, both
analytically and numerically, that the surface multifractal
exponents $\tau_q^{\rm s}(b,p)$ are not only different from their
bulk counterparts, $\tau_q^{\rm b}(b)$, but also depend on an
additional parameter $p$ characterizing the reflection of the
particle at the boundary. This peculiarity of the PRBM model is
intimately related to the existence of the line of fixed points
(labelled by $b$) in the bulk model. Indeed, the freedom in the choice
of the amplitude of the boundary ``hopping with reflection'' term is
of the same origin as the freedom in the amplitude of the power-law
hopping in the bulk. 

We close on a somewhat speculative note. The existence of a truly
marginal coupling (implying a line of fixed points) is not unique
for the PRBM model. In particular, the 2D Dirac fermions in a random
vector potential \cite{MF-Dirac} share this feature. Furthermore, it
was conjectured recently \cite{CFT-IQH-Zirnbauer,CFT-IQH-Tsvelik}
that the quantum Hall transition might be described by a particular
point on a line of fixed points in a related model. Based on our
results, it is natural to ask whether the emergence of an additional
parameter $p$ governing the boundary criticality is a general
property of critical theories with a truly marginal coupling. The
work in this direction is currently underway \cite{unpub}.

\section{Acknowledgments}
\label{s6}


The work of AM, RN, FE, and ADM was supported by the Center for
Functional Nanostructures and the Schwerpunktprogramm
``Quanten-Hall-Systeme'' of the Deutsche Forschungsgemeinschaft. The
work of IAG and ARS was supported by the NSF MRSEC Program under
DMR-0213745, the NSF Career Award DMR-0448820, the Sloan Research
Fellowship from Alfred P Sloan Foundation and the Research
Innovation Award from Research Corporation. ARS, IAG, and ADM
acknowledge hospitality of the Kavli Institute for Theoretical
Physics at Santa Barbara and support by the National Science
Foundation under Grant No.\ PHY99-07949.




\begin{thebibliography}{99}


\bibitem[$*$]{byline1} Also at Petersburg Nuclear Physics
Institute, 188300 St.~Petersburg, Russia.

\bibitem{anderson58} P. W. Anderson,
Phys. Rev. {\bf 109}, 1492 (1958).

\bibitem{Wegner80} F. Wegner, Z. Phys. B {\bf 36}, 209 (1980).

\bibitem{Castellani-Peliti-86} C. Castellani and L. Peliti,
J. Phys. A {\bf 19}, L429 (1986).

\bibitem{janssen} M.~Janssen, Phys. Rep. {\bf 295}, 1 (1998).

\bibitem{review}  A. D. Mirlin, Phys. Rep. {\bf 326}, 259
(2000).

\bibitem{duplantier91} B.~Duplantier and A.~W.~W.~Ludwig, Phys. Rev.
Lett. {\bf 66}, 247 (1991).

\bibitem{akl} B. L.~Altshuler, V. E.~Kravtsov, and I. V.~Lerner,
Sov. Phys. JETP {\bf 64}, 1352 (1986).

\bibitem{MF-Anderson-numerics} M.~Schreiber and H.~Grussbach,
Phys. Rev. Lett. {\bf 67}, 607 (1991); A.~Mildenberger, F.~Evers,
and A. D.~Mirlin, Phys. Rev. B {\bf 66}, 033109 (2002).

\bibitem{MF-weak} V. I. Fal'ko and K. B. Efetov, Europhys. Lett.
{\bf 32}, 627 (1995); Phys. Rev. B {\bf 52}, 17\,413 (1995).

\bibitem{MF-Dirac} A.~W.~W.~Ludwig, M.~P.~A.~Fisher, 
R.~Shankar, and G.~Grinstein, Phys. Rev. B {\bf 50}, 7526 (1994); 
C.~Mudry, C.~Chamon, and X.-G.Wen, Nucl. Phys. B {\bf 466}, 383 (1996); 
H.~E.~Castillo, C.~Chamon, E.~Fradkin, P.~M.~Goldbart, and C.~Mudry, 
Phys. Rev. B {\bf 56}, 10\,668 (1997); J.-S. Caux, N.~Taniguchi, 
and A.~M.~Tsvelik, Nucl. Phys. B {\bf 525}, 671 (1998);
Phys. Rev. Lett. {\bf 80}, 1276 (1998); J.-S. Caux, Phys. Rev. Lett.
{\bf 81}, 4196 (1998).

\bibitem{MF-symplectic} S.~N.~Evangelou, Physica A {\bf 167}, 199 (1990);
J.~T.~Chalker, G.~J.~Daniell, S.~N.~Evangelou, and I.~H.~Nahm,
   J. Phys.: Cond. Matter {\bf 5}, 485 (1993);
L.~Schweitzer, J. Phys. C {\bf 7}, L281 (1995);
T.~Kawarabayashi and T.~Ohtsuki, Phys. Rev. B  {\bf 53}, 6975, (1996);
K.~Yakubo and M.~Ono, Phys. Rev. B {\bf 58}, 9767 (1998);
A.~Mildenberger and F.~Evers, Phys. Rev. B {\bf 75}, 041303(R) (2007).

\bibitem{MF-IQH} W.~Pook and M.~Janssen, Z. Phys. B {\bf 82}, 295
(1991); B.~Huckestein, B.~Kramer, and L.~Schweitzer, Surf. Sci. {\bf
263}, 125 (1992); B.~Huckestein, Rev. Mod. Phys. {\bf 67}, 357
(1995); R.~Klesse and M.~Metzler, Europhys. Lett, {\bf 32}, 229
(1995); Int. J. Mod. Phys. C {\bf 10}, 577 (1999); M.~Janssen, 
M.~Metzler, and M.~R.~Zirnbauer, Phys. Rev. B {\bf 59}, 15\,836 (1999);
F.~Evers, A.~Mildenberger, and A.~D.~Mirlin, Phys. Rev. B {\bf 64},
241303(R) (2001).

\bibitem{Mirlin03} A.~D.~Mirlin, F.~Evers, and A.~Mildenberger, 
J. Phys. A {\bf 36}, 3255 (2003).

\bibitem{MF-RBM} A.~D.~Mirlin and F.~Evers, Phys. Rev. B {\bf 62},
7920 (2000).

\bibitem{subramaniam06} A.~R.~Subramaniam, I.~A.~Gruzberg,
A.~W.~W.~Ludwig, F.~Evers, A.~Mildenberger, and A.~D.~Mirlin,
Phys. Rev. Lett. {\bf 96}, 126802 (2006).

\bibitem{sqhe-unpub} A.~R.~Subramaniam, I.~A.~Gruzberg,
and  A.~W.~W.~Ludwig, to be published.

\bibitem{obuse06} H.~Obuse, A.~R.~Subramaniam, A.~Furusaki,
I.~A.~Gruzberg, and A.~W.~W.~Ludwig, cond-mat/0609161, accepted
in Phys. Rev. Lett.

\bibitem{prbm} A. D. Mirlin, Y. V. Fyodorov, F.-M. Dittes, J. Quezada,
  and T. H. Seligman, Phys. Rev. E 54, 3221 (1996).

\bibitem{levitov90} L.~S.~Levitov, Phys. Rev. Lett. {\bf 64}, 547
(1990); B.~L.~Altshuler and L.~S.~Levitov, Phys. Rep. {\bf 288}, 487
(1997).

\bibitem{prbm-papers} 
V.~E.~Kravtsov and K.~A.~Muttalib, Phys. Rev. Lett. {\bf 79}, 1913 (1997);
V.~E.~Kravtsov and A.~M.~Tsvelik, Phys. Rev. B {\bf 62}, 9888 (2000);
E.~Cuevas, M.~Ortuno, V.~Gasparian, and A.~Perez-Garrido, Phys. Rev. Lett.
{\bf 88}, 016401 (2001); I.~Varga, Phys. Rev. B {\bf 66}, 094201
(2002); E.~Cuevas, Phys. Rev. B {\bf 68}, 024206 (2003); {\it ibid.}
{\bf 68}, 184206 (2003); {\it ibid.} {\bf 71}, 024205 (2005);
O.~Yevtushenko and V.~E.~Kravtsov, J. Phys. A: Math. Phys. {\bf 36},
8265 (2003); Phys. Rev. E {\bf 69}, 026104
(2004); A.~M.~Garcia-Garcia and K.~Takahashi, Nucl.Phys. B {\bf 700},
361 (2004); J.~A.~Mendez-Bermudez and T.~Kottos, Phys. Rev. B {\bf
72}, 064108 (2005); A.~M.~Garcia-Garcia, Phys.Rev. E {\bf 73} 026213
(2006); V.~E.~Kravtsov, O.~Yevtushenko, and E.~Cuevas, J. Phys. A:
Math. Gen. {\bf 39}, 2021 (2006). J.~A.~Mendez-Bermudez and I. Varga,
Phys. Rev. B {\bf 74}, 125114 (2006).

\bibitem{fyodorov94} Y.~V.~Fyodorov and A.~D.~Mirlin,
  Int. J. Mod. Phys. B {\bf 8}, 3795 (1994).

\bibitem{efetov-book} For a detailed exposition of the supersymmetry
  method the reader is referred to the book, K.~B.~Efetov, {\it
    Supersymetry in disorder and chaos} (Cambridge University Press,
  1997).

\bibitem{levy-flights} M.~F.~Schlesinger, G.~M.~Zaslavsky, and
  J.~Klafter, Nature {\bf 363}, 31 (1993); R.~Metzler and J.~Klafter,
  Phys. Rep. {\bf 339}, 1 (2000).

\bibitem{zumofen95}  G.~Zumofen and J.~Klafter, Phys. Rev. E {\bf 51},
  2805 (1995).

\bibitem{levy-p1} R.~Metzler and J.~Klafter, Physica A {\bf 278},
  107 (2000); N.~Krepysheva, L.~Di~Pietro, and M.-C.~N\'eel,
Phys. Rev. E {\bf 73}, 021104 (2006).

\bibitem{note1} Strictly speaking, Eqs.~(\ref{e26}), (\ref{e25}) are
valid for all $q>1/2$ in the limit $b\to 0$. For a finite (but
small) $b$, Eq.~(\ref{e26}) breaks down in a narrow interval of $q$
above $1/2$, namely for $q-1/2 \lesssim 1/\ln b^{-1}$. Indeed, the
evolution equation (\ref{e23}) assumes that resonances are rare,
i.e., that the angle $\theta$ describing the resonant mixture is
large compared to its typical, nonresonant, value $\sim b$. On the
other hand, when $q$ approaches 1/2, the integral in Eq.~(\ref{e25})
converges at $\theta\sim\exp[-1/(q-1/2)]$. Comparing this with $b$,
we get the above restriction on the validity of Eq.~(\ref{e26}).


\bibitem{mirlin06} A.~D.~Mirlin, Y.~V.~Fyodorov, A.~Mildenberger, and
F.~Evers, Phys. Rev. Lett. {\bf 97}, 046803 (2006).

\bibitem{CFT-IQH-Zirnbauer} M. R. Zirnbauer, hep-th/9905054.

\bibitem{CFT-IQH-Tsvelik} M.~J.~Bhaseen, I.~I.~Kogan, O.~A.~Soloviev,
N.~Taniguchi, and A.~M.~Tsvelik, Nucl. Phys.~B {\bf 580}, 688 (2000).

\bibitem{unpub} A. R. Subramaniam {\it et al.},
unpublished.

\end{thebibliography}
\end{document}